\documentclass[aps,prl,reprint,groupedaddress,showpacs]{revtex4-1}

  \usepackage{graphicx}
  \usepackage{dcolumn}
  \usepackage{bm}
  \usepackage[table]{xcolor}

\newcommand{\VLDA}{V_c^\textrm{\scriptsize LDA}}
\newcommand{\gLDA}{\gamma^\textrm{\scriptsize LDA}}
\newcommand{\EPs}{E_\textrm{\scriptsize Ps}}
\newcommand{\TF}{_\textrm{\scriptsize TF}}

\begin{document}

\title{Proposed parameter-free model for interpreting the measured positron annihilation spectra of materials using a generalized gradient approximation}

\author{Bernardo Barbiellini}
\email{B.Amidei@neu.edu}
\affiliation{
Department of Physics, Northeastern University, Boston, Massachusetts 02115, USA}

\author{Jan Kuriplach}
\email{Jan.Kuriplach@mff.cuni.cz}
\affiliation{
Department of Low Temperature Physics, Faculty of Mathematics and Physics, 
Charles University, V Hole\v{s}ovi\v{c}k\'ach 2, CZ-180\,00 Prague, Czech Republic}


\begin{abstract}
Positron annihilation spectroscopy is often used to analyze the local electronic structure of materials of technological interest.
Reliable theoretical tools are crucial to interpret the measured spectra.
Here, we propose a parameter-free gradient correction scheme for a local-density approximation obtained from high quality quantum Monte Carlo data.
The results of our calculations compare favorably with positron affinity and lifetime measurements opening new avenues for highly precise and advanced positron characterization of materials. 
\end{abstract}

\pacs{78.70.Bj, 71.60.+z, 71.15.Mb}

\maketitle


The positron upon annihilation with its anti-particle, the electron, yields unique information about the electronic structure of bulk materials \cite{review_tdft,Tuomisto} and nanostructures \cite{Eijt}.
The electron-positron density functional theory (DFT) \cite{tdft} is used in order to obtain 
precise knowledge of the positron wave function and its overlap with the electron orbitals. 
The powerful combination of positron annihilation spectroscopy (PAS) and DFT calculations provides 
a highly accurate method for advanced characterization of materials \cite{bba_review}.

Within the DFT framework, the generalized gradient approximation (GGA) method to describe electron-positron correlation effects in solids has shown a systematic improvement over the local density approximation (LDA) for positron affinities and annihilation characteristics \cite{gga1,gga2,campillo,ggaboro}.
Until now, a dimensional analysis has been used to determine the form of the lowest-order gradient correction with a semi-empirical coefficient $\alpha$.
So far, the pragmatic approach has been to fit $\alpha$ to large databases of positron lifetimes.
Recently, both the LDA and the GGA \cite{gga4,gga5} have been improved on the basis of new quantum Monte Carlo (QMC) data for the electron-positron correlation problem in a homogenous electron gas (HEG) \cite{qmc}.

However, one could claim that such good fits may be in some cases accidental \cite{stachowiak}.
Moreover, the present gradient corrections may also lead to some unphysical effects in the electron-positron correlation potential near the nuclei; namely, its too large oscillations due to the shell structure of core electrons.
Therefore, here we propose to improve the GGA by extracting and deducing the $\alpha$ parameter from more fundamental physical principles. 
This more reliable derivation of $\alpha$ also reveals a gentle dependence of the local density reducing the gradient correction near the nuclei. 
Thus, $\alpha$ becomes a function of the local density as well.

In the case of the positron immersed in an electron
gas the Coulomb attraction produces a cusp in the electron
density at the positron site.
The correlation potential describing the positron perturbation represents the electronic polarization due to the positron screening and can be obtained via the Hellmann-Feynman theorem using coupling-constant integration as follows \cite{hodges73}
\begin{equation}
V_{c}(\bm{r}) = -\int_{0}^{1}\!\text{d}Z \int\!\text{d}^3\!\bm{R}\,\,
\frac{\rho(\bm{R})\,[g(\bm{r},\bm{R},Z){-}1]}{|\bm{r-R}|} ~, 
\end{equation}
where $\rho(\bm{R})\,[g(\bm{r},\bm{R},Z){-}1]$ is the screening cloud density around a positive particle with charge $Z$ ($g(\bm{r},\bm{R},Z)$ is the particle-electron pair distribution function).
The effect of the density gradient on the correlation energy can be deduced 
from the distortion of the polarization cloud due to this gradient.
For this purpose, one can use the dynamical structure factor 
$S(\bm{q},\omega)$ \cite{Lindhard,Abbamonte} of the HEG to show that in the high density limit the lowest order gradient correction is proportional to the parameter
$\epsilon=(|\nabla \ln \rho|/q\TF)^2$ (which depends on the ratio of
the Thomas-Fermi length $\lambda\TF = 1/q\TF$ and the inhomogeneity length $1/|\nabla \ln \rho|$). 
This correction is given by the expression 
\begin{equation}
\Delta V_{c}(\bm{r})= \beta \frac{\epsilon(\bm{r})}{16} ~,
\label{eqcg} 
\end{equation}
where the constant $\beta=0.066725$ Hartree is linked to the coefficient of the term $q^2$ in the density response function wave vector expansion.
The coefficient $\beta$ has been calculated by Ma and Brueckner \cite{MB} and has been used by various authors \cite{LM1,LM2,ePBEXC}.
Eq. (\ref{eqcg}) is in fact similar to that used to compute the correlation energy \cite{ePBEXC} for an electron gas with slowly varying density 
\footnote{The difference originates from the extra screening in the presence of two types of carriers which produces a rescaling by a factor of 1/4 in the gradient correction coefficient \cite{MB}.
}.

In order to interpolate to the case of rapid density variations (i.e. large $\epsilon$), 
we use the formula 
\begin{equation}
V_c = \VLDA \exp(-\alpha\:\!\epsilon/3) ~,
\end{equation}
from Ref. \cite{gga1} [see Eq. (7) there]. 
This formula is based on the scaling relation for the correlation potential, as derived by Nieminen and Hodges \cite{scalingrel}.
But $\alpha$ is now a function of the local density (and thereby position).
When we identify the first order expansion in $\epsilon$ with the Ma and Brueckner's result shown above, we find that 
\begin{equation}
\alpha(\bm{r}) = - \frac{3}{16}\, \frac{\beta}{\VLDA(\bm{r})} ~.
\label{PFalpha}
\end{equation}
The quantity $\alpha$ remains a gentle function of the density in the valence electron region and at low density it becomes very close to $0.05$ \footnote{Since $V_c$ approaches 1/4 Hartree for small densities, $\alpha \approx 3\beta/4 \doteq 0.05$ if Hartree units are used for $\beta$.} -- a value found earlier within the empirical GGA \cite{gga4,gga5}. 
Interestingly, $\alpha$ happens also to be of the same order as the fraction $Z_c$ of an electron displaced in electron-electron correlation effects which is typically of the order of 1/20 of the electron charge \cite{coulson,bba89}.

Like the potential $V_c$, the positron annihilation rate depends on electron-positron correlation effects and must be enhanced over the independent particle model.
The electron-positron enhancement theory \cite{Makkonen14}
has some features in common with the interaction between a core hole and the 
conduction electrons treated both in X-ray emission \cite{carbotte68} and in resonant
inelastic X-ray scattering \cite{hancock14}. 
We can relate the correlation energy to the annihilation rate by using the scaling relation \cite{scalingrel}. 
Therefore, one obtains an electron-positron enhancement annihilation factor $\gamma$ given by
\begin{equation}
\gamma-1 = (\gLDA-1)\,\exp(-\alpha\:\!\epsilon) ~,
\end{equation}

The enhancement term $\gamma$ is used to calculate the total positron annihilation rate or the inverse lifetime $1/\tau$, which is expressed through the simple relation \cite{bba_review}
\begin{equation}
\frac{1}{\tau}=
\pi r_0^2 c\, \int \text{d}^3\bm{r}\,\gamma(\bm{r})\, \rho(\bm{r}) \,
|\psi_+(\bm{r})|^2 \,,
\end{equation}
where $r_0$ is the classical radius of the electron, $c$ is the speed of light
and $\psi_+(\bm{r})$ is the ground state positron wave function.


In this work, we have used the same accurate computational method described in Refs.
\cite{gga4, gga5}. 
Electronic structure calculations for selected materials were carried out using the self-consistent WIEN2k code \cite{wien2k}, which imposes no shape restrictions for the electron density and the potential, while the positron wave function and energy were obtained using a Schr\"odinger equation solver based on a finite difference method.
The exchange-correlation potential for the electrons contains gradient corrections within the scheme proposed by Perdew, Burke and Ernzerhof \cite{ePBEXC}
except in the case of the 4$d$ and 5$d$ elemental metals since some of their calculated properties (e.g. the lattice constant) become inappropriate when gradient corrections are used \cite{bmj90}. 
GGA corrections introduce cusps in the electron potential, negligible in the LDA, which reflect the atomic shell structure \cite{bmj90}.
Numerical parameters of the WIEN2k code as well as of the positron solver were tested and optimized in order to obtain calculated positron lifetimes within a precision of 0.1 ps and positron affinities within 0.01 eV.
Here we consider only systems in which the positron density is approaching zero in the limit of an infinite crystal.


\begin{table}[htb]
\vspace*{-3mm}
\caption{Positron affinities (in eV) calculated according to various approaches:
GC = original gradient correction with the Arponen and Pajanne potential \cite{AP} ($\alpha=0.22$),
DB = Drummond {\em et al.} \cite{qmc}, 
DG = gradient correction with DB ($\alpha=0.05$),
PF = parameter-free gradient correction with DB (varying $\alpha$).
The last column gives experimental values taken from Refs. \cite{Gidley88,Jibaly91,weiss}. 
The exceptions are C and Si (see Ref. \cite{gga4} and references therein) and 
MgO (Ref. \cite{nagashima}).
In the case of MgO an upper limit is given (see the text).
}
\begin{ruledtabular}
\begin{tabular}[t]{@{\ }l l r r r r l}
System & Structure & GC & DB & DG & PF & Exp. \\    
\hline\\[-2mm]
\multicolumn{7}{c}{Elements} \\
Li    & bcc       & $-7.31$ & $-7.02$ & $-6.95$ & $-6.96$    \\
C     & diamond   & $-1.33$ & $-2.40$ & $-1.87$ & $-1.93$ & $-2.0$ \\
Na    & bcc       & $-7.18$ & $-6.89$ & $-6.80$ & $-6.81$          \\
Al    & fcc       & $-4.21$ & $-4.04$ & $-4.00$ & $-4.01$ & $-4.1$ \\
Si    & diamond   & $-6.29$ & $-6.47$ & $-6.33$ & $-6.35$ & $-6.2$ \\
Fe    & bcc       & $-3.40$ & $-3.76$ & $-3.62$ & $-3.67$ & $-3.3$ \\
Cu    & fcc       & $-3.76$ & $-4.23$ & $-4.05$ & $-4.11$ & $-4.3$ \\
Nb    & bcc       & $-3.61$ & $-3.75$ & $-3.65$ & $-3.68$ & $-3.8$ \\
Ce    & fcc, $\alpha$-Ce& $-4.11$ & $-4.16$ & $-4.07$ & $-4.09$   \\
Ce    & fcc, $\gamma$-Ce& $-5.34$ & $-5.31$ & $-5.23$ & $-5.25$   \\
W     & bcc       & $-1.72$ & $-1.91$ & $-1.82$ & $-1.85$ & $-1.9$ \\
Pt    & fcc       & $-3.31$ & $-3.77$ & $-3.61$ & $-3.67$ & $-3.8$ \\
\multicolumn{7}{c}{Compounds} \\
MgO   & rock salt & $-5.56$ & $-6.46$ & $-6.17$ & $-6.25$ & $-5.2$ \\
Cu$_2$O & cuprite &  $-5.88$ & $-6.42$ & $-6.21$ & $-6.26$   \\
CeO$_2$ & fluorite & $-6.55$ & $-7.40$ & $-7.12$ & $-7.18$   \\
YBa$_2$Cu$_3$O$_6$  & tetragonal   &  $-6.11$ & $-6.65$ & $-6.42$ & $-6.46$ \\
YBa$_2$Cu$_3$O$_7$  & orthorhombic &  $-6.02$ & $-6.78$ & $-6.52$ & $-6.58$ \\
PrBa$_2$Cu$_3$O$_7$ & orthorhombic&  $-5.81$ & $-6.57$ & $-6.30$ & $-6.36$ \\ 
\end{tabular}
\end{ruledtabular}
\label{tabA}
\vspace*{-1mm}
\end{table}

\begin{table}[htb]
\caption{Positron lifetimes (in ps) calculated according to various approaches explained in the caption of Table \ref{tabA}.
The last column gives experimental values discussed in Refs. \cite{gga4,gga5}.
The last experimental values for cuprates are extracted from Refs. \cite{shukla,bba91}.
}
\begin{ruledtabular}
\begin{tabular}[t]{@{\ }l l r r r r l}
System & Structure & GC & DB & DG & PF & Exp. \\
\hline\\[-2mm]
\multicolumn{7}{c}{Elements} \\
Li    & bcc       & 283.2 & 303.8 & 316.2 & 313.5 & 291  \\
C     & diamond   & 102.8 &  94.6 &  98.9 &  97.7 & 98+  \\
Na    & bcc       & 337.7 & 343.0 & 364.4 & 360.5 & 338  \\
Al    & fcc       & 154.2 & 161.0 & 164.1 & 163.0 & 160+ \\
Si    & diamond   & 222.7 & 208.1 & 217.3 & 215.9 & 216+ \\
Fe    & bcc       & 109.6 & 102.1 & 106.5 & 104.7 & 105+\\
Cu    & fcc       & 120.0 & 107.4 & 113.3 & 110.9 & 110+ \\
Nb    & bcc       & 123.4 & 120.9 & 124.3 & 123.1 & 120+ \\
Ce    & fcc, $\alpha$-Ce& 169.5 & 165.0 & 170.5 & 169.0 & 233 \\
Ce    & fcc, $\gamma$-Ce& 196.8 & 194.1 & 200.6 & 198.9 & 235 \\
W     & bcc       & 102.7 & 100.6 & 103.4 & 102.3 & 105  \\
Pt    & fcc       & 105.2 &  97.4 & 101.3 &  99.8 & 99+  \\
\multicolumn{7}{c}{Compounds} \\
MgO    & rock salt & 146.2 & 119.0 & 128.5 & 125.4 & 130 \\
Cu$_2$O & cuprite  & 177.4 & 147.3 & 158.4 & 154.8 & $\sim$174 \\
CeO$_2$ & fluorite & 173.7 & 138.2 & 149.1 & 146.0 & $<$187 \\
YBa$_2$Cu$_3$O$_6$    & tetragonal   & 224.5 & 175.4 & 190.8 & 186.5 & $\sim$190 \\
YBa$_2$Cu$_3$O$_7$    & orthorhombic & 179.2 & 142.4 & 154.0 & 150.5 & $\sim$165 \\
PrBa$_2$Cu$_3$O$_7$    & orthorhombic& 180.4 & 143.4 & 155.0 & 151.6 & $\sim$165 \\ 
\end{tabular}
\end{ruledtabular}
\label{tabL}
\vspace*{-1mm}
\end{table}

DFT provides an excellent description of the Si electronic structure both in the solid and liquid phases \cite{Okada12}. 
It is therefore natural to start our tests of the parameter-free GGA positron potential in Si. A meaningful observable to check is the  positron affinity $A$ defined as the sum of the electron and positron chemical potentials. 
In the case of a semiconductor, the electron chemical potential is taken from the position of the top of the valence band.  
Recently, Cassidy {\em et al.} \cite{Cassidy11} have  shown that the temperature invariant time of flight (TOF) component for Ps emitted from the surface of $p$-doped Si(100) has a kinetic energy equal to $0.6$ eV. 
This TOF feature is explained by a bulk positron picking up a valence band electron just beneath the surface to form Ps with a kinetic energy of $K=\EPs+A=0.6$ eV. 
Therefore, the experimental affinity for Si can be deduced to be  $A=-6.2$ eV. 
When we use the GGA for both the electron and positron potentials, we find a theoretical value $A=-6.35$ eV, which is in excellent agreement with the value measured by Cassidy {\em et al.} while the corresponding LDA value shows a clear tendency to overestimate the magnitude of $A$. 
This LDA problem can be traced back to the screening effects. 
In the GGA, the value of $A$ agrees with the experiment by reducing the screening charge.  Calculated positron affinities within LDA and GGA against the corresponding experimental values for different materials are shown in Table \ref{tabA}. 
The trends follow those of Si, nevertheless the experimental values of $A$ are often of earlier date and not always reliable. 
The corresponding positron lifetimes are presented in Table \ref{tabL}. 
Clearly, the trends of the parameter-free GGA are very similar to the empirical GGA \cite{gga4,gga5}.
In particular, one of the best result is given by Al which was problematic in the original GGA scheme \cite{gga1}.
Positron lifetime measurements in Li and Na were performed before the advent of reliable spectrometers and fitting procedures, as discussed in detail in Ref. \cite{gga4}, and may be affected by significant errors.

\begin{figure}
 \begin{center}
 \includegraphics[width=8.7cm,clip]{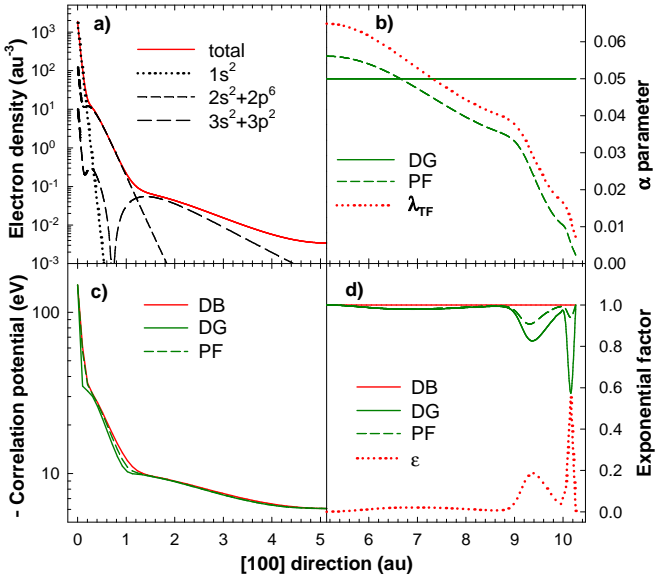}
  \end{center}
  \caption{(Color online) One-dimensional profiles of (a) the electron density (including atomic orbitals), (b) the $\alpha$ parameter (and the Thomas-Fermi length $\lambda\TF$), (c) the positron correlation potential, and (d) the exponential factor $\exp(-\alpha\epsilon/3)$ (and $\epsilon$ parameter) along the [100] direction in Si for LDA (DB), the empirical (DG) and the parameter-free (PF) GGA approaches. 
Si atoms are located at 0 and 10.26 au along [100].
$\lambda\TF$ and $\epsilon$ are shown for the purpose of observing correlations with corresponding quantities (the scales of $\lambda\TF$ and $\epsilon$ are different than those for $\alpha$ and $\exp(-\alpha\epsilon/3)$, respectively).} 
  \label{fig1}
  \end{figure}

\begin{figure}
 \begin{center}
 \includegraphics[width=8.7cm,clip]{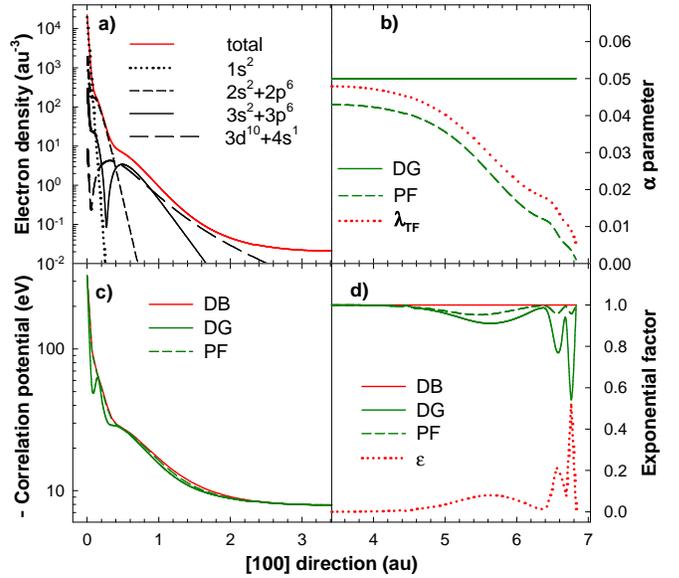}
  \end{center}
  \caption{(Color online) 
  One dimensional profiles for Cu as explained in the caption of Fig. \ref{fig1}.
  Cu atoms are located at 0 and 6.83 au along the [100] direction.
  } 
  \label{fig2}
  \end{figure}

However, in the present scheme the positron has a slightly larger overlap with the core electrons as illustrated in Figs. \ref{fig1} and \ref{fig2} for Si and Cu, respectively. Some noticeable jumps of $\epsilon$ shown in Fig.~\ref{fig1} (d)
and Fig.~\ref{fig2} (d) result in unphysically large local changes in the 
empirical GGA correlation potential depicted in Fig. \ref{fig1} (c) and Fig. \ref{fig2} (c). 
These problems are now cured by the variation of the function $\alpha$  
in space illustrated by Fig.~\ref{fig1} (b) and Fig.~\ref{fig2} (b). 
Interestingly, $\alpha$ given by Eq. (\ref{PFalpha}) seems to vary almost like the Thomas-Fermi length $\lambda\TF$ and becomes very small close to the nuclei. 
Therefore, the cusps in the parameter free GGA correlation potential become more damped because of the reduction of the screening length in the core region.
This effect is further documented by $\exp(-\alpha\epsilon/3)$ factor plots (Fig. \ref{fig1} (d) and Fig. \ref{fig2} (d)) which define the reduction of the correlation potential in the core region.
The $\exp(-\alpha\epsilon/3)$ factor anticorrelates with the $\epsilon$ parameter;
i.e. a large inhomogeneity corresponds to a small exponential factor.
The variation of $\alpha$ in the core region should also improve the description of high-momentum annihilation spectra observed in coincidence Doppler broadening spectroscopy \cite{alatalo2,mijnarends98} and in angular correlation measurements \cite{Laverock}.

The positron annihilation lifetime (PAL) provides a way to detect very small amounts of vacancy-defects in crystalline materials. 
Since thermalized positrons are trapped by vacancies before annihilating with electrons, their lifetime increases with respect the bulk values given the low electron density at the vacancy. 
For this reason, PAL has been widely used to characterize doped semiconducting samples of silicon and other technological relevant materials \cite{Tuomisto}. 
As shown by Table \ref{tabL} the positron bulk lifetime of Si is very well described by the present theory. 
Therefore, deviations from the theoretical lifetime indicate the presence of imperfections in the sample. 
In a post-silicon electronics era, engineered doping of oxide electronics, which is similar to conventional doping in semiconductor technology, offers much greater functionality including electronic control of redox chemistry with applications to batteries, photovoltaics and catalysis. In particular, a well characterized material is MgO, which is a simple binary oxide with rock-salt structure. 
In MgO, a magnetic moment can arise from the unpaired 2p electrons at an oxygen site surrounding a cation vacancy with each nearest neighbor oxygen carrying a magnetic moment \cite{araujo10}. 
This magnetic property can be fine tuned to optimize spintronics devices. 
Concerning PAL studies, Tanaka {\em et al}. \cite{MgOdoping} have shown that MgO lifetime is  significantly affected by Ga doping, which results in the creation of Mg vacancies. 
However, when the number of Mg vacancies decreases the lifetime converges to the bulk value 130 ps \cite{MgOPAS}, which is in reasonable agreement with the present theory.
A reliable experimental TOF study of MgO \cite{nagashima} reports a Ps emission peak energy of $2.6$ eV. 
Since Ps is already formed in the bulk of MgO, the kinetic energy is given in this case by 
$K = \EPs + A - E_B + E_G$, where $E_B$ is the Ps binding energy inside the MgO matrix and $E_G=7.8$ eV is the energy gap of MgO.  
Using our calculated affinity, we deduce that $E_B=5.75$ eV, which is consistent with typical values of Ps binding energy in the bulk \cite{kolkata}. 
In fact, this value must be smaller than $\EPs$ because of screening effects in the bulk.

Ceria \cite{jarlborg14} is another oxide which has attracted considerable interest because of its applications in solid oxide fuel cells.  It can be noted that by removing all the oxygen atoms, one recovers the fcc structure of Ce. Experimentally, positron seems only to detect the $\gamma$ phase of Ce because of its stronger affinity with respect to the $\alpha$ phase. Interestingly, the experimental ceria lifetime 189 ps \cite{ceriaLT1} appears to be much closer to theoretical value of $\gamma$-Ce rather than ceria. A possible reason  for this discrepancy is that real samples can always contain patches of $\gamma$-Ce which strongly attract the positron because of their higher positron affinity. In this context, we should keep in mind that oxygen is very mobile in ceria.

As an example of advanced characterization, we now show that positron annihilation spectroscopy can be useful to understand the role of oxygen-related defects in high temperature superconductivity \cite{icpa16}. 
In practice, by comparing the experimental lifetimes \cite{shukla} to an accurate theory it is possible to deduce that positrons are trapped at oxygen vacancies in the superconducting compound YBa$_2$Cu$_3$O$_{7{-}\delta}$ while this trapping becomes negligible in the non-superconducting compound where Y has been replaced by Pr. 
When positrons become completely delocalized for temperatures higher than 400 K, the lifetime becomes almost identical in the YBa$_2$Cu$_3$O$_7$ and PrBa$_2$Cu$_3$O$_7$ compounds in agreement with our calculations reported in Table \ref{tabL}. 
Moreover, the calculated lifetime in the tetragonal YBa$_2$Cu$_3$O$_6$ lattice is 36 ps longer than in the orthorhombic YBa$_2$Cu$_3$O$_7$. 
Such difference is consistent with experiments \cite{bba91}. 
Curiously, the calculated positron affinity seems to indicate that Ps is emitted with about $0.15$ eV higher kinetic energy from YBa$_2$Cu$_3$O$_6$ and PrBa$_2$Cu$_3$O$_7$ than from YBa$_2$Cu$_3$O$_7$. 
Nevertheless, since the present DFT calculations fail in describing the insulating phase of YBa$_2$Cu$_3$O$_6$ and PrBa$_2$Cu$_3$O$_7$ we should take the positron affinity calculated values for these two compounds with caution.  

In conclusion, 
we have demonstrated that the parameter-free GGA truly provides a simple, yet accurate step beyond LDA.
It is also reassuring that the most reliable electron-positron LDA parametrization (based on the QMC simulations) combined with the parameter free gradient correction gives the best results compared with any of the older LDA potentials.
Further studies combining the present approach with well-converged momentum densities calculations \cite{ernsting14} are needed to check if first principle methods can soon improve the agreement over empirical approaches \cite{Laverock}.

\begin{acknowledgments}
We acknowledge fruitful discussions with A.P. Mills and Y. Nagashima.
B.B. is supported by the U.S. Department of Energy (USDOE) Contract No. DE-FG0207ER46352 and has benefited for computer time from Northeastern University's Advanced Scientific Computation Center (ASCC) and USDOE’s NERSC supercomputing center.
J.K. acknowledges the support by the IT4Innovations Centre of Excellence project (CZ.1.05/1.1.00/02.0070), funded by the European Regional Development Fund and the national budget of the Czech Republic via the Research and Development for Innovations Operational Programme, as well as Czech Ministry of Education, Youth and Sports via the project Large Research, Development and Innovations Infrastructures (LM2011033).
\end{acknowledgments}

\bibliography{ggam}

  \end{document}